\documentclass[12pt,a4paper]{article}
\usepackage[english]{babel}
\usepackage{amsmath,amsthm,amssymb,epsfig,latexsym}
\usepackage{color}
\usepackage{ulem}
\usepackage[numbers,square,sort&compress]{natbib}

\newcommand{\blu}{\color{blue}}

\newcommand{\so}{\scriptscriptstyle \rm I}
\newcommand{\st}{\scriptscriptstyle \rm I\hspace{-1pt}I}
\newcommand{\sth}{\scriptscriptstyle \rm I\hspace{-1pt}I\hspace{-1pt}I}

\newcommand{\tz}{\tilde z}

\newcommand{\bet}{\bar\eta}

\newcommand{\be}[1]{\begin{equation}\label{#1}}
\newcommand{\ba}[1]{\begin{multline}\label{#1}}
\newcommand{\ee}{\end{equation}}
\newcommand{\ea}{\end{multline}}

\theoremstyle{plain}
\newtheorem{thm}{Theorem}[section]
\newtheorem{prop}{Proposition}[section]
\newtheorem{lemma}{Lemma}[section]

\theoremstyle{definition}
\newtheorem{Def}{Definition}[section]

\newtheorem{rem}{Remark}[section]
\def\qed{\hfill\nobreak\hbox{$\square$}\par\medbreak}
 \makeatletter
 \@addtoreset{equation}{section}
 \makeatother
 
\newcommand{\bea}{\begin{eqnarray}}
\newcommand{\eea}{\end{eqnarray}}

\def\F{{\mathcal{F}}}

\def\BB{{\mathbb{B}}}
\def\CC{{\mathbb{C}}}
\newcommand{\ZZ}{{\mathbb Z}}

\def\FF{{\rm F}}

\def\rvac{|0\rangle}

\def\EE{{\rm E}}
\def\FF{{\rm F}}

\def\tFF{\tilde{\rm F}}

\def\Ee{\mathbf{e}}

\def\r#1{\eqref{#1}}

\def\sk#1{\left(#1\right)}
\def\Bal{\mathcal{B}}

\def\Pepm{{P}^\pm_e}

\def\Pfpm{{P}_f^\pm}
\def\Pfp{{P}^+_f}
\def\Pfm{{P}^-_f}

\def\fgo{\mathfrak{f}}

\def\BF{\mathbb{F}}
\def\KK{\mathbb{K}}

\def\kk{\chi}
\def\twis{\chi}

\def\bet{\bar\eta}
\def\DYBn{\mathcal{D}Y(\mathfrak{o}_{2n+1})}

\def\Defun#1{\delta_{#1}}

\def\gfu{\gamma}

\begin{document}
\pagestyle{empty}
\setcounter{page}{1}

\vspace{12pt}

\begin{center}
\begin{LARGE}
{\bf Algebraic Bethe ansatz for\\ $\mathfrak{o}_{2n+1}$-invariant  integrable models}
\end{LARGE}

\vspace{40pt}

\begin{large}
{A.~Liashyk${}^{a,b}$,
S.~Z.~Pakuliak${}^{c,d}$  \footnote{
a.liashyk@gmail.com, stanislav.pakuliak@jinr.ru}}
\end{large}

\vspace{10mm}

${}^a$ {\it Skolkovo Institute of Science and Technology, Moscow, Russia}

\vspace{2mm}

${}^b$ {\it National Research University Higher School of Economics, Russia}

\vspace{2mm}

${}^c$ {\it Moscow Institute of Physics and Technology,  Dolgoprudny, Moscow reg., Russia}

\vspace{2mm}

${}^d$ {\it Laboratory of Theoretical Physics, JINR,  Dubna, Moscow reg., Russia}

\vspace{2mm}

\end{center}


\vspace{4mm}

\begin{abstract}
A class of  $\mathfrak{o}_{2n+1}$-invariant quantum integrable models is investigated in the framework of algebraic Bethe ansatz method. 
A construction of the  $\mathfrak{o}_{2n+1}$-invariant  Bethe vector is proposed 
in terms of the Drinfeld currents for the double of Yangian $\mathcal{D}Y(\mathfrak{o}_{2n + 1})$. 
Action of the monodromy matrix entries onto off-shell Bethe vectors for these models is calculated. Recursion relations for these 
vectors were obtained. The action formulas can be used 
 to investigate  structure of the
scalar products of Bethe vectors in $\mathfrak{o}_{2n+1}$-invariant  models.

\end{abstract}

\pagestyle{plain}

\section{Introduction}

Quantum integrable systems where the algebraic Bethe ansatz may be applied are described by the
monodromy matrices that satisfy quadratic relations defined by some $R$-matrix.
Main goal of the algebraic Bethe ansatz is to construct state vectors or Bethe vectors
for integrable models from matrix elements of the corresponding monodromies.  
Eigenvalue property of the Bethe vectors with respect  to the transfer matrix
 uses these quadratic commutation relations between monodromy matrix entries
 and leads to the Bethe equations
on the parameters of Bethe vectors.

In this paper, we will consider quantum integrable systems whose monodromy matrices satisfy 
the commutation relations with 
the $\mathfrak{o}_{2n + 1}$-invariant $R$-matrix proposed in \cite{ZZ79,Resh85} and solved in \cite{Resh91}. 
This $R$-matrix is associated with an 
infinite-dimensional algebra, that is
double of Yangian $\mathcal{D}Y(\mathfrak{o}_{2n + 1})$ \cite{Dr88, D88}. 
Generating series  of generators of this
algebra can be assembled into a matrix or $T$-operator that will satisfy the same quadratic
commutation relations as the monodromy matrix of $\mathfrak{o}_{2n + 1}$-invariant
integrable system. The coincidence of these commutation relations allows
to construct state vectors in terms of generators of the algebra $\mathcal {D}Y(\mathfrak {o}_{2n + 1})$
and use the properties of this algebra to study the space of states in
$\mathfrak{o}_{2n + 1}$-invariant integrable system.

The advantages of this approach to the algebraic Bethe ansatz are that doubles of Yangians and
quantum affine algebras, in addition to the $R$-matrix formulation, also have a 'new' realization
in terms of formal series or 'currents' proposed by \cite{D88}. For quantum affine
algebra $U_q(\widehat{\mathfrak{gl}}_n)$ the equivalence of the two formulations was proved
in \cite{DF93}, and for similar infinite-dimensional algebras associated with
algebras of the series $B$, $C$ and $D$ analogous equivalences were recently proved in \cite{JLY18, JLM18, JLM19, JLM19a}.

The main tools in proving the equivalence of various realizations
of quantum affine algebras and doubles of Yangians are the so-called {\it Gaussian coordinates}
of $T$-operators. These objects turned out to be very useful in research of the space of states 
in quantum integrable models associated
with trigonometric and rational deformations of affine algebra
$\widehat{\mathfrak{gl}}_n$ and their super-symmetric extensions
\cite{KhP-Kyoto, PRS13, HutLPRS17a}. Gaussian coordinates may be related to the   currents
in the 'new' realization of the Yangians and quantum affine algebras \cite{D88} using
projections to the intersections of the different types Borel subalgebras. The Bethe vectors itself
can be constructed as projections from the products of currents \cite{EKhP07}.

Until recently the description of the space of states in the quantum integrable models 
using the current realization of the corresponding infinite-dimensional algebras was available only 
for the models associated with serie $A$ algebras.
Recent results published in  \cite{JLY18,JLM18,JLM19,JLM19a} open
the possibility to develop tools for the investigation of the 
quantum integrable models associated with infinite-dimensional algebras for
$B$, $C$ and $D$ series. 
In particular, the relations between Gaussian coordinates for the Yangian doubles of 
$B$, $C$ and $D$ series and projections of the corresponding currents were established in \cite{LP20}.
Note also the papers \cite{KK19,vidas, vidas2}, where the different methods of investigations
of the orthogonal and symplectic quantum integrable models were developed.

The present paper is a continuation of the research started in  \cite{LPRS19a} where
a description of the off-shell Bethe vectors for the integrable models associated 
with the algebra $\mathfrak{o}_3$ was obtained. 
In many cases, to investigate the space of states of quantum integrable model one does not need
the explicit formulas for the Bethe vectors in terms of monodromy entries. It is sufficient to explore 
the explicit formulas of the actions of the monodromy matrix elements onto off-shell Bethe vectors.  
In \cite{HutLPRS20}  such action formulas were presented in case of the 
supersymmetric $\mathfrak{gl}(m|n)$-invariant integrable models.
In this paper we obtain the explicit formulas for such an action for 
the Bethe vectors in the $\mathfrak{o}_{2n+1}$-invariant quantum integrable  models.  

The paper is composed as follows. Section~\ref{RTTalg} is devoted to definition of the 
class of $\mathfrak{o}_{2n+1}$-invariant quantum integrable  models based on $R$-matrix 
found by A.B.~Zamolodchikov and Al.B.~Zamolodchikov. 
Main results of our research is presented in the section~\ref{MMact}
together with demonstration of the different reductions of the main action formula to already 
known cases.  Last section~\ref{BVvsC} contains the proof of two  technical statements 
necessary for the calculation of the action formulas. These proofs use  the projection method formulated 
for this case in \cite{LP20}. 

\section{$\mathfrak{o}_{2n+1}$-invariant $R$-matrix and $RTT$ algebra}\label{RTTalg}

Let $N=2n+1$ for positive integer $n\geq1$ and $\Ee_{i,j}$ are $N\times N$ matrices
with the only nonzero entry equal to $1$ at the intersection of the $i$-th row and $j$-th column. 
We will use  integers to number matrix entries of the operators 
in End$(\CC^N)$: $-n\leq i,j\leq n$.
Let $\mathbf{P}$ (permutation operator) and $\mathbf{Q}$ be  operators  acting in
$\CC^N\otimes\CC^N$
\begin{equation*}\label{PQop}
\mathbf{P}=\sum_{i,j=-n}^n\Ee_{i,j}\otimes\Ee_{j,i},\quad 
\mathbf{Q}=\sum_{i,j=-n}^n\Ee_{i,j}\otimes\Ee_{-i,-j} \,.
\end{equation*}

We denote by $R(u,v)$  $\mathfrak{o}_{2n+1}$-invariant $R$-matrix    \cite{ZZ79}
\begin{equation}\label{Rmat}
  R(u,v) = \mathbf{I}\otimes\mathbf{I} + \frac{c\, \mathbf{P}}{u-v} - \frac{c\,\mathbf{Q}}{u-v+c\kappa_n}\,,
\end{equation}
where $\mathbf{I}=\sum_{i=-n}^{n}\Ee_{i,i}$ is the identity operator  in $\CC^{N}$, 
 $c$ is a constant, $u$ and $v$ are arbitrary complex parameters called spectral parameters
and 
\begin{equation*}\label{kappan}
\kappa=N/2-1=n-1/2. 
\end{equation*}

Due to the properties 
\begin{equation*}\label{PQpro}
\mathbf{P}\  \mathbf{Q}=\mathbf{Q}\ \mathbf{P}=\mathbf{Q},\quad
\mathbf{P}^2= \mathbf{I}\otimes\mathbf{I},\quad  \mathbf{Q}^2=N\mathbf{Q}
\end{equation*}
this $R$-matrix obeys  the Yang-Baxter equation 
\begin{equation*}\label{YB}
R_{1,2}(u_1,u_2)\cdot R_{1,3}(u_1,u_3)\cdot R_{2,3}(u_2,u_3) =
R_{2,3}(u_2,u_3)\cdot R_{1,3}(u_1,u_3)\cdot R_{1,2}(u_1,u_2) 
\end{equation*}
in  $\CC^N\otimes\CC^N\otimes\CC^N$  
(here subscripts of $R$ denote  spaces $\CC^N$ in which $R$-matrix acts) and 
satisfies the unitarity condition
\begin{equation}\label{Runi}
R(u,v)\ R(v,u)=\sk{1-\frac{c^2}{(u-v)^2}}\mathbf{I}\otimes\mathbf{I}\,.
\end{equation}

Let $T(u)$ be an operator valued $N\times N$ matrix satisfying the quadratic commutation relations 
\begin{equation}\label{rtt}
  R(u,v) \left( T(u)\otimes\mathbf{I} \right) \left( \mathbf{I}\otimes T(v) \right) =
  \left( \mathbf{I}\otimes T(v) \right) \left( T(u)\otimes\mathbf{I} \right) R(u,v)\,.
\end{equation}
These commutation relations can be written in terms of the matrix entries  $T_{i,j}(u)$
\begin{equation*}\label{Tent}
T(u)=\sum_{i,j=-n}^n \Ee_{i,j}\ T_{i,j}(u)\,,
\end{equation*}
\begin{equation}\label{rrt2}
\begin{split}
 & \left[ T_{i,j}(u), T_{k,l}(v) \right] =  \frac{c}{u-v}\left( T_{k,j}(v)T_{i,l}(u) - T_{k,j}(u) T_{i,l}(v) \right)+\\
  &\qquad+\frac{c}{u-v+c\kappa}\left(\delta_{k,-i}\sum_{p=-n}^{n} T_{p,j}(u)T_{-p,l}(v)-
  \delta_{l,-j}\sum_{p=-n}^{n} T_{k,-p}(v)T_{i,p}(u)\right)\,.
  \end{split}
\end{equation}

As a direct consequence of the commutation relations  \r{rrt2} 
the matrix elements  
should satisfy the relations \cite{JLM18}
\begin{equation}\label{qu-rel}
\sum_{m=-n}^{n} T_{-m,-i}(u-c\kappa) T_{m,j}(u)=
\sum_{m=-n}^{n} T_{i,m}(u) T_{-j,-m}(u-c\kappa)= z(u) \delta_{i,j}\,,
\end{equation}
where $z(u)$ is a central elements  in the $RTT$ algebra \r{rrt2}. 
In what follows we set this central element to 1. 
 The quadratic relations \r{qu-rel} imposes nontrivial relations between entries 
  $T_{i,j}(u)$.

\subsection{$\mathfrak{o}_{2n+1}$-invariant  integrable models}\label{o2n1}

We denote the $RTT$ algebra \r{rtt} as $\Bal$ algebra. The algebraically 
dependent  elements of this algebra are operators 
$T_{i,j}[\ell]$, $\ell\geq 0$, $-n\leq i,j\leq n$ gathered into generating series
\begin{equation}\label{depen}
T_{i,j}(u)=\delta_{ij}+\sum_{\ell\geq0} T_{i,j}[\ell](u/c)^{-\ell-1}. 
\end{equation}
The zero modes operators $T_{i,j}[0]\equiv T_{i,j}$ will play a special role below. 

The generating series \r{depen} 
can be used to construct  
$\mathfrak{o}_{2n+1}$-invariant  integrable models. Let $\mathcal{H}$
be a Hilbert space of states of such a model, which can be considered as representation space for the algebra $\Bal$. 
In order to treat this model by the algebraic Bethe ansatz method, the physical space of 
the model $\mathcal{H}$ must include a special 
vector or reference state $\rvac$ such that 
\begin{equation}\label{vac}
\begin{split}
T_{i,j}(u)\rvac &=0, \quad \quad {-n\leq j<i\leq n},\\
 T_{i,i}(u)\rvac &=\lambda_i(u)\rvac,  \quad -n\leq i\leq n ,
\end{split}
\end{equation}
where $\lambda_i(u)$ characterize the concrete model. 
They are free functional parameters modulo the relations \r{rest} 
which follow from \r{qu-rel}.
As a result of the commutation relations \r{rtt} the transfer matrix 
\begin{equation}\label{trans}
\mathcal{T}(z)=\sum_{i=-n}^{n} T_{i,i}(z)
\end{equation}
generates a commuting set of integrals of the model: $[\mathcal{T}(u), \mathcal{T}(v)]=0$.

Let $t^i_\ell\in\CC$, $i=0,1,\ldots,n-1$, $\ell=1,\ldots,r_i$ be  generic complex parameters,  where 
the positive integers 
$r_i\geq 0$ count the number of parameters of the type $i$. These numbers 
are cardinalities of the sets $\bar t^i$: $r_i=\#\bar t^i$ and if some $r_i=0$ then the corresponding 
set $\bar t^i$ is empty: $\bar t^i=\varnothing$. We will gather these parameters into sets
\begin{equation}\label{sets}
\bar t=\{\bar t^0,\ldots,\bar t^{n-1}\},\quad \bar t^i=\{t^i_1,\ldots,t^i_{r_i}\},\quad i=0,1,\ldots,n-1.
\end{equation}
For $\ell=1,\ldots,r_i$, the notation $\bar t^i_\ell=\{\bar t^i\setminus t^i_\ell\}$ stands for the set $\bar t^i$ with the parameter 
$t^i_\ell$ omitted.  Then, $\bar t^i_\ell$ has cardinality $\#\bar t^i_\ell=r_i-1$.
We will call the parameters $t^i_\ell$ as the {\it Bethe parameters}.

In what follows we will need the following rational functions  
\begin{equation}\label{fun-g}
g(u,v)=\frac{c}{u-v},\quad f(u,v)=\frac{u-v+c}{u-v},\quad h(u,v)=\frac{f(u,v)}{g(u,v)}=
\frac{u-v+c}{c}\,.
\end{equation}
To simplify presentation of our results we introduce the set of functions
\begin{equation}\label{fun-go}
 f_s(u,v)=\begin{cases}
& \fgo(u,v)=\displaystyle{\frac{u-v+c/2}{u-v}},\quad s=0\,,\\
& f(u,v),\quad s=1,\ldots,n-1.
 \end{cases}
\end{equation}
There is an equality
\begin{equation*}\label{fgo-rel}
f(u,v)=\fgo(u+c/2,v)\fgo(u,v)\,.
\end{equation*}

We will use a shorthand notation for the products of  functions of one or two variables.
For example, 
\begin{equation*}\label{not}
\lambda_k(\bar u)=\prod_{u_i\in\bar u} \lambda_k(u_i),\qquad \fgo(u,\bar v)=\prod_{v_j\in\bar v}\fgo(u,v_j),\qquad
f(\bar u,\bar v)=\prod_{u_i\in\bar u}\prod_{v_j\in\bar v}f(u_i,v_j).
\end{equation*}
If any set in these formulas is empty, then the corresponding product is equal to 1.

The algebraic Bethe ansatz allows to describe the physical states of the model in terms of Bethe vectors.
The vectors $\BB(\bar t)$ are called {\it on-shell} Bethe vectors if the set $\bar t$ of the Bethe parameters satisfies the so-called Bethe equations
\begin{equation}\label{BE}
\alpha_s(t^s_\ell)= 
\frac{f_s(t^s_\ell,\bar t^s_\ell)}{f_s(\bar t^s_\ell,t^s_\ell)}
\frac{f(\bar t^{s+1},t^s_\ell)}{f(t^s_\ell,\bar t^{s-1})},\quad  \ell=1,\ldots,r_s,\quad s=0,1,\ldots,n-1,
\end{equation}
where we assume $\bar t^{-1}=\bar t^{n}=\varnothing$ and define  functions 
\begin{equation}\label{alpdef}
 \alpha_s(z)=\frac{\lambda_{s}(z)}{\lambda_{s+1}(z)}.
\end{equation}

On-shell Bethe vectors are eigenstates of the transfer matrix 
\begin{equation}\label{ABA}
\mathcal{T}(z)\BB(\bar t)=\tau(z;\bar t)\BB(\bar t).
\end{equation}

To describe the eigenvalue $\tau(z;\bar t)$   it is convenient to introduce the notation 
\begin{equation}\label{zs}
z_s=z-c\sk{s-\frac{1}{2}},\quad s=0,1,\ldots,n-1,n\,.
\end{equation}
Note that $z_{n}=z-c\kappa$. The eigenvalue $\tau(z;\bar t)$ in \r{ABA}
is equal to 
\begin{equation}\label{eigBA}
\begin{split}
\tau(z;\bar t)&= \lambda_{0}(z) f(\bar t^0,z_0) f(z,\bar t^0)\ +\\
&\quad+\sum_{s=1}^n \Big(\lambda _{s}(z) f(\bar t^{s},z) f(z,\bar t^{s-1})
+ \lambda _{-s}(z) f(\bar t^{s-1},z_{s-1}) f(z_{s},\bar t^{s})\Big)\,,
\end{split}
\end{equation}
where 
\begin{equation}\label{rest}
\lambda_{-j}(z)=\frac{1}{\lambda_{n}(z_{n})}\prod_{s=j}^{n-1}
\frac{\lambda_{s+1}(z_s)}{\lambda_{s}(z_s)}
= \frac{1}{\lambda_{j}(z_{j})}\prod_{s=j+1}^{n}
\frac{\lambda_{s}(z_{s-1})}{\lambda_{s}(z_{s})} 
\end{equation}
for $j=0,1,\ldots,n$.

The Bethe equations \r{BE} can be obtained as vanishing of the residues of the eigenvalue $\tau(z;\bar t)$
\r{eigBA} at the values $z=t^s_\ell$ or  $z_s=t^s_\ell$.
If the Bethe parameters $\bar t$ are free, such vectors are called {\it off-shell} Bethe vectors. 
There exist different methods to describe the off-shell Bethe vectors $\BB(\bar t)\in\mathcal{H}$  in terms of
 polynomials of the non-commuting monodromy matrix elements $T_{i,j}(u)$, $i < j$  acting onto 
reference vector $\rvac$. But it appears that in order to calculate physically interesting quantities in such integrable models, one does not need 
to have fully explicit formulas for these vectors. In many cases it is sufficient to know explicit formulas for the action of the monodromy 
matrix elements $T_{i,j}(z)$ onto off-shell Bethe vectors. One can prove that the set of these vectors is closed under this action, which means 
that the product  $T_{i,j}(z)\cdot\BB(\bar t)$ maybe written as a linear combination of the vectors of the same structure. 

The search of the explicit formulas for such an action in $\mathfrak{o}_{2n+1}$-invariant integrable models 
is the main goal of this paper. 

\section{Monodromy matrix elements action}\label{MMact}

In this section we present the main result of this paper.  
Besides collections of the Bethe parameters $\bar t$ defined by \r{sets} 
we introduce the collection of sets $\bar w$ 
  \begin{equation}\label{wsets1}
\bar w=\{\bar w^0,\bar w^1,\ldots,\bar w^{n-1}\},\quad
\bar w^s=\{ t^{\,s}_1,\ldots,t^{\,s}_{r_s},z,z_s\},\quad 
\end{equation}
with $z_s$ defined by \r{zs}. 
Let $\bar w^n=\{z,z_n\}$ be an auxiliary set of the cardinality 2.

The action of the monodromy matrix 
element $T_{i,j}(z)$, $-n\leq i,j\leq n$ onto off-shell Bethe vector $\BB(\bar t)$ is given 
by the sum over partitions
of the sets $\{\bar w^s_{\so},\bar w^s_{\st},\bar w^s_{\sth}\}\vdash \bar w^s$,
$s=0,1,\ldots,n$  such that 
cardinalities of the sets  $\bar w^s_{\so}$ and $\bar w^s_{\sth}$ are 
\begin{equation}\label{card}
\begin{split}
\#\bar w^s_{\so}&=\Theta(i+s)+\Theta(i-s-1)\,,\\
\#\bar w^s_{\sth}&=\Theta(s-j)+\Theta(-j-s-1)\,.
\end{split}
\end{equation}
Here $\Theta(p)$ is a Heaviside step function defined for integers $p\in\ZZ$ 
\begin{equation*}
  \Theta(p)  = 
 	\begin{cases}
 	 	1,    \quad  p\geq 0,  \\ 
 	 	0,    \quad  p<0.
 	\end{cases}	
\end{equation*}

The cardinalities of the sets 
$\bar w^s_{\so}$ and $\bar w^s_{\sth}$ may be equal to 0, 1 or 2 depending on the interrelations between 
integers $i$, $j$ and $s$.  Let us define 
\begin{equation}\label{sign}
\sigma_i=2\Theta(i-1)-1. 
\end{equation}
It is clear that $\sigma_i=-1$ for $i\leq 0$ and $\sigma_i=1$ for $i>0$.
Note also, that according to \r{card} $\#\bar w^n_{\so}=\#\bar w^n_{\sth}=1$
for all values of the indices $-n\leq i,j\leq n$.

Let us define the rational functions of two variables
\begin{equation}\label{t-fun}
\gfu_s(u,v)=\begin{cases}
\begin{split}     \displaystyle{\fgo(u,v)} &= \displaystyle{\frac{u - v + c/2}{u - v}}, && s=0,\\
  \displaystyle{\frac{f(u,v)}{h(u,v)h(v,u)}}&=\displaystyle{\frac{c^2}{(u-v)(v-u+c)}} , && s=1,\ldots,n-1.
\end{split}
 \end{cases}
\end{equation}

The main result of this paper may be formulated as  following 
\begin{thm}\label{mainprop}
The action of the monodromy matrix element $T_{i,j}(z)$, $-n\leq i,j\leq n$ 
onto off-shell Bethe vector $\BB(\bar t)$ is given by the explicit formula
\begin{equation}\label{mainac}
\begin{split}
&T_{i,j}(z)\cdot \BB(\bar t)=-\frac{\sigma_{i}\sigma_{-j}\ \lambda_{n}(z)\ g(z_1,\bar t^0)}{\kappa\ h(z,\bar t^0)}
\sum_{\rm part}\  \BB(\bar w_{\st})\prod_{s=0}^{n-1}
\alpha_s(\bar w^s_{\sth}) \\
&\quad\times\prod_{s=0}^{n-1}
\gfu_s(\bar w^s_{\so},\bar w^s_{\st})
\gfu_s(\bar w^s_{\st},\bar w^s_{\sth})
\gfu_s(\bar w^s_{\so},\bar w^s_{\sth})\ 
\frac{h(\bar w^{s+1}_{\st},\bar w^{s}_{\so})
h(\bar w^{s+1}_{\sth},\bar w^{s}_{\so})
h(\bar w^{s+1}_{\sth},\bar w^{s}_{\st})}
{g(\bar w^{s+1}_{\so},\bar w^{s}_{\st})
g(\bar w^{s+1}_{\so},\bar w^{s}_{\sth})
g(\bar w^{s+1}_{\st},\bar w^{s}_{\sth})}.
\end{split}
\end{equation}
Here we use notation $\bar w_{\st}=\{\bar w^0_{\st},\bar w^1_{\st},\ldots,\bar w^{n-1}_{\st}\}$. 
The sets $\bar w^s$ for $s=0,1,\ldots,n$ are divided into subsets
 $\{\bar w^s_{\so},\bar w^s_{\st},\bar w^s_{\sth}\}\vdash \bar w^s$  and summation 
 in \r{mainac} goes over these partitions with cardinalities described by \r{card}.
\end{thm}

Recall that functions $\alpha_s(z)$ are defined by \r{alpdef}.

\begin{rem}\label{reading}
It follows from the action formula that effectively the sum over partitions of the set 
$\bar w^n$ reduces only to one term when $\bar w^{n}_{\so}=\{z_n\}$ and $\bar w^{n}_{\sth}=\{z\}$.
Indeed, let us assume that 
$\bar w^{n}_{\so}=\{z\}$ since $\#\bar w^n_{\so}=1$. Then the product $g(z,\bar w^{n-1}_{\st})^{-1}
g(z,\bar w^{n-1}_{\sth})^{-1}$ in denominator of the second line of \r{mainac} yields
$z\in\bar w^{n-1}_{\so}$ otherwise we got zero term. Considering 
other factors $g(\bar w^{s+1}_{\so},\bar w^{s}_{\st})^{-1} g(\bar w^{s+1}_{\so},\bar w^{s}_{\sth})^{-1}$ 
for $s=0,n-2$ we prove analogously that in order to have non-zero terms 
$z$ should be in the set $\bar w^s_{\so}$ for all $s$. Analogously, let us assume that 
$\bar w^{n}_{\sth}=\{z_n\}$. The product $h(z_n,\bar w^{n-1}_{\so})
h(z_n,\bar w^{n-1}_{\st})$ yields $z_{n-1}\in\bar w^{n-1}_{\sth}$ since otherwise we got 
zero contribution because $h(z_n,z_{n-1})=0$. Continuing we prove that assumption 
$\bar w^{n}_{\sth}=\{z_n\}$ yields that  $z_{s}\in\bar w^{s}_{\sth}$ for all $s$. 
But \r{mainac} has a factor $\fgo(\bar w^0_{\so},\bar w^0_{\sth})\sim\fgo(z,z_0)\equiv 0$. 
This proves that our initial assumptions that  $\bar w^{n}_{\so}=\{z\}$ and $\bar w^{n}_{\sth}=\{z_n\}$ 
lead to the zero contribution into sum over partitions. 
\end{rem}

\begin{rem}\label{embed} (\textit{Reduction to $\mathfrak{o}_{2n-1}$-invariant Bethe vectors}.) 
By the same arguments as in the previous remark we may observe that the action formula 
respect the hierarchical embedding of the $\mathfrak{o}_{2n-1}$-invariant $RTT$ algebra 
into $\mathfrak{o}_{2n+1}$-invariant $RTT$ algebra described in lemma~(3.6) of the paper 
\cite{JLM18}. It means, that if $\bar t^{n-1}=\varnothing$ then we can prove that 
$\bar w^{n-1}_{\so}=\{z_{n-1}\}$ and $\bar w^{n-1}_{\sth}=\{z\}$ and that
the action formula \r{mainac}
for the values of the indices $-n+1\leq i,j\leq n-1$ becomes the action formula for 
the $\mathfrak{o}_{2n-1}$-invariant integrable model. 
\end{rem}

\subsection{Zero-modes method}

To prove the statement of the  theorem~\ref{mainprop} we will use so called {\it zero-modes method}. To describe 
it we modify slightly the definition of the algebra $\Bal$ given in the section~\ref{o2n1}. 
Let 
\begin{equation*}\label{Kdiag}
\KK={\rm diag}(\kk_{-n},\ldots,\kk_{-1},1,\kk_{1},\ldots,\kk_{n})
\end{equation*}
 be a diagonal matrix with non-zero entries $\kk_i\in\CC$. Sometimes we will use in the formulas below the notation $\kk_{0}=1$.
The diagonal matrix $\KK$ satisfies the relation 
\begin{equation}\label{ZM6}
R(u,v)\cdot \KK_1\otimes \KK_2 = \KK_1\otimes \KK_2 \cdot R(u,v)
\end{equation}
if
\begin{equation}\label{zmrel2}
    \twis_{-i}=\twis_{i}^{-1},\quad i=1,\ldots,n.
\end{equation}
So it has only $n$ independent parameters $\twis_i$ for $i=1,\ldots,n$.  
Due to \r{ZM6}, the monodromy matrix $T^\KK(u)=\KK\cdot T(u)$ satisfies the same commutation relations  \r{rrt2}. 
The only difference will be that instead of the expansion \r{depen} we will 
have\footnote{Recall that we denoted zero modes $T_{i,j}[0]$ as $T_{i,j}$ for $-n\leq i,j\leq n$.}
\begin{equation}\label{texp}
 T^\KK_{i,j}(u) = \kk_i \delta_{ij}  + \frac{c}{u}  \,  T_{i,j} + O( u^{-2}).
\end{equation}

The commutation relations \r{rrt2} yield the commutation of the zero modes 
and monodromy matrix elements\footnote{In what follows we remove 
the superscript in the notation of the monodromy matrix elements $T^\KK_{i,j}(u)$ always assuming 
expansion \r{texp}.}
\begin{equation}\label{ZM7}
  \Big[ T_{i,j}(u), T_{k,l} \Big] = \delta_{i,l}\,\kk_i\, T_{k,j}(u) -\delta_{k,j}\, \kk_j\, T_{i,l}(u) - 
 \delta_{i,-k}\, \kk_l\, T_{-l,j}(u)+ \delta_{-l,j}\,\kk_k\, T_{i,-k}(u). 
\end{equation}

Proof of the theorem~\ref{mainprop} is based on the commutation 
relations \r{ZM7} and two propositions.

\begin{prop}\label{ZMp}
The action of the zero mode operators $T_{j,i}$ onto off-shell Bethe vectors $\BB(\bar t)$ 
for $0\leq i<j\leq n$ is given by the equality 
\begin{equation}\label{ZMpart}
\begin{split}
   &T_{j,i}\cdot \BB( \bar t )  =   \sum_{\rm part}   \BB(\bar t_{\st})
    \prod_{s=1}^{n-1}\frac{1}{g(\bar t^{s}_{\st},\bar t^{s-1}_{\so})g(\bar t^{s}_{\so},\bar t^{s-1}_{\st})
    g(\bar t^{s}_{\so},\bar t^{s-1}_{\so})}
        \frac{1}{h(\bar t^s_{\st},\bar t^s_{\so})h(\bar t^s_{\so},\bar t^s_{\st})}
     \\
      &\qquad\times
     \Bigg(  \twis_{j}  
      \prod_{s=i}^{j-1}
      \frac{ \alpha_{s}(\bar t^{s}_{\so}) f(\bar t^{s}_{\so},\bar t^{s-1})f_s(\bar t^{s}_{\st},\bar t^{s}_{\so})}
      { h(\bar t^{s}_{\so},\bar t^{s-1}_{\so})}
        -
       \twis_{i} \prod_{s=i}^{j-1}
       \frac{  f(\bar t^{\,s+1},\bar t^{\,s}_{\so}) f_s(\bar t^{\,s}_{\so},\bar t^{\,s}_{\st}) }
       {h(\bar t^{s+1}_{\so},\bar t^s_{\so})} \Bigg) ,
       \end{split}
\end{equation}
where the functions $\alpha_s(t)$ are defined by \r{BE} and the sum goes over partitions 
$\{\bar t^s_{\so},\bar t^s_{\st}\}\vdash \bar t^s$ with cardinalities $\#\bar t^s_{\so}=1$ for 
$s=i,\ldots,j-1$ and $\#\bar t^s_{\so}=0$ for other
$s$. 
\end{prop}

\begin{rem}\label{zermodbe}  If $\kk_j=1$ and the Bethe parameters $\bar t$  satisfy the Bethe equations \r{BE}
the on-shell Bethe vectors become highest weight vectors for the algebra $\mathfrak{o}_{2n+1}$.  Literally, it means that in this case $T_{j,i}\cdot \BB( \bar t ) = 0$.
\end{rem}

Note that the product in the first line of \r{ZMpart} is going effectively from $s=i$ to $s=j$.

 \begin{prop}\label{achimo}
  The action of monodromy matrix element $T_{-n,n}(z)$ onto off-shell Bethe vector 
  $\BB(\bar t)$ \r{beth1} is regular and given by the relation
    \begin{equation}\label{dr3}
    T_{-n,n}(z)\cdot \BB(\bar t)   =
  -\kappa\ \frac{g(z_1,\bar t^{\,0}) h(z,\bar t^{\,n-1})}{
    \textstyle{h(z,\bar t^{\,0}) g(z_n,\bar t^{\,n-1}) }} \  \lambda_{n} (z)\ 
 \BB (\bar w)\,,
  \end{equation}
  and the collection of sets  $\bar w$ is defined by \r{wsets1}.
   \end{prop}

Proofs of the propositions~\ref{ZMp} and \ref{achimo}
will be given  in  section~\ref{BVvsC} using identification of $RTT$ algebra 
$\Bal$ with the standard Borel subalgebra of the Yangian double $\DYBn$ \cite{Dr88}.
This infinite-dimensional algebra was
investigated  in \cite{JLY18,LP20} and we will use certain projections onto intersections 
of the different types Borel subalgebras studied in \cite{EKhP07} for the 
quantum affine algebras.

As a direct consequence of \r{ZM7} the zero-modes operators obey the relations
\begin{equation}\label{ZM777}
  \Big[ T_{i,j}, T_{k,l} \Big] = \delta_{i,l}\,\kk_i\,T_{k,j} - \delta_{k,j}\,\kk_j\, T_{i,l} - 
 \delta_{i,-k}\, \kk_l\, T_{-l,j}+\delta_{-l,j}\, \kk_k\, T_{i,-k}\,.
\end{equation}

As well as for the generating series $T_{i,j}(u)$   
the quadratic relation \r{qu-rel}  impose several relations to the zero-modes operators. 
Substituting expansion \r{texp} into \r{qu-rel} and equating terms at $u^0$  and 
$u^{-1}$ we obtain \eqref{zmrel2} and ($-n\leq i,j\leq n$)
\begin{equation}\label{zmrel}
 \twis_{-j}   T_{i,j}    + \twis_{i}  T_{-j,-i} = 0\,.
\end{equation}

It follows from  \r{zmrel2} and \r{zmrel} that 
\begin{equation}\label{zmrel3}
    T_{-j,-i}=-\twis_i^{-1}\twis_j^{-1}T_{i,j},\quad -n\leq i,j\leq n
\end{equation}
and this equality for $j=-i$ implies that 
\begin{equation*}\label{zmrel4}
    T_{i,-i}=-T_{i,-i}=0,\quad  -n\leq i\leq n\,.
\end{equation*}
Equality \r{zmrel3} and commutation relations \r{ZM777} allow to express all zero-modes 
through the algebraically independent set of generators $T_{i,i}$, $T_{i-1,i}$, $T_{i,i-1}$ for $i=1,\ldots,n$.

Let us sketch the proof of the theorem~\ref{mainprop} using  zero-modes method. 
We start from \r{dr3} and commutation relation  
\begin{equation}\label{comz1}
 [T_{-n,n}(z) , T_{n,i} ] = - \twis_{n} T_{-n,i}(z) - \twis_{i} T_{-i,n} (z)
\end{equation}
for $0\leq i\leq n-1$ to obtain from \r{ZMpart} the action formula for the operators 
$T_{-i,n} (z)$. This can be achieved by applying the equality \r{comz1} to the off-shell 
Bethe vector $\BB(\bar t)$ and equating the coefficients at $\twis_i$ in left and right 
hand side of this equality. This can be done because $\twis_i$ are independent parameters.

Next we start from the action of the entry $T_{0,n}(z)$  and applying the commutation relations
\begin{equation*}\label{comz2}
 [T_{0,n}(z) , T_{i,0} ] =  T_{i,n}(z) -\delta_{i,n} \twis_{n} T_{0,0} (z)
\end{equation*}
to $\BB(\bar t)$ for $1\leq i\leq n$ we obtain the action of the rest entries $T_{i,n}(z)$ from 
the last column of monodromy matrix $T(z)$.

At the next step we explore already calculated actions of the entries 
$T_{i,n}(z)$, $-n\leq i\leq n$   and the commutation 
relation 
\begin{equation*}\label{comz3}
 [T_{i,n}(z) , T_{n,j} ] =  - \twis_{n} T_{i,j}(z) - 
 \delta_{ i,-n} \twis_{j} T_{-j,n}(z) +
\delta_{i,j} \twis_{i} T_{n,n}(z)
\end{equation*}
to obtain the action formulas of the entries $T_{i,j}(z)$ for all $i$ and $0\leq j\leq n-1$.

Finally, we start from the action of the entries $T_{i,0}(z)$ for all $i$ and use the commutation relation 
\begin{equation*}\label{comz4}
 [T_{i,0}(z) , T_{j,0} ] =  \twis_{j} T_{i,-j}(z) - 
 \delta_{ i,-j}  T_{0, 0}(z) +
\delta_{i,0} \twis_{i} T_{j,0}(z)
\end{equation*} 
to obtain action formulas for remaining entries $T_{i,-j}(z)$, $-n\leq i\leq n$ and $1\leq j\leq n$.
Performing these calculations we can always equate the terms at the independent 
parameters $\twis_j$, ${ 0}\leq j\leq n$.\qed

\subsection{Recurrence relations}

The action formulas \r{mainac} allows to obtain the recurrence relations for the off-shell Bethe vectors 
in $\mathfrak{o}_{2n+1}$-invariant integrable models. For $\mathfrak{gl}(m|n)$-invariant 
supersymmetric  models such recurrence relations were obtained in \cite{HLPRS17b}. 

\begin{prop}\label{recrel}
For $n\geq2$ one has following recurrence relations for the off-shell Bethe vector $\BB(\bar t)$
\begin{equation}\label{rr1}
\begin{split}
 &  \BB(\bar t^{\,0},\ldots,
 \bar t^{\,n-2},\{\bar t^{\,n-1},z\}) = \frac{1}{h(z,\bar t^{\,n-1}) \lambda_{n}(z)} \\ 
 &\quad  \times \sum_{i=-n}^{n-1} 
    \sum_{{\rm part}} \frac  {\sigma_{i+1}T_{i ,n}(z)\cdot \BB(\bar t_{\st} )} { h(\bar t^{\,n-1}_{\st},z)g(z,\bar t^{\,n-2}_{\st})}\ 
     \prod_{s=0}^{n-1}  \gamma_s(\bar t^{\,s}_{\so},\bar t^{\,s}_{\st}) 
     \prod_{s=1}^{n-1}
    \frac{h(\bar t^{\,s}_{\st},\bar t^{\,s-1}_{\so})}{g(\bar t^{\,s}_{\so},\bar t^{\,s-1}_{\st})}   ,
    \end{split}
\end{equation}
where sum in \r{rr1} goes over partitions $\{ \bar t^{\,s}_{\so},\bar t^{\,s}_{\st}\}\vdash \bar t^{\,s}$ 
with different cardinalities of $\bar t^{\,s}_{\so}$ for different terms depending on $i$ for $s\leq n-2$ as
\begin{equation}\label{rr2}
\#\bar t^s_{\so}=\begin{cases}0,\quad 0\leq s\leq i-1\,,\\ 1,\quad i\leq s\leq n-2
\end{cases}
\end{equation}
for $0\leq i\leq n-1$, 
\begin{equation*}\label{rr3}
\#\bar t^s_{\so}=\begin{cases}2,\quad 0\leq s\leq -i-1\,,\\ 1,\quad -i\leq s\leq n-2
\end{cases}
\end{equation*}
for $-n+1\leq i\leq -1$ and  for $s = n-1$
\begin{equation}\label{rr4}
\#\bar t^{n-1}_{\so}=\begin{cases}0,\quad -n+1\leq i\leq n-1\,,\\ 1,\quad i=-n\,.
\end{cases}
\end{equation}
\end{prop}

Proof of this proposition can be performed in a similar way as in \cite{HLPRS17b}. We have to use 
the action formulas \r{mainac} in the right hand side of \r{rr1} and prove that the right hand side identically 
coincide with the left hand side in this equality. The recurrence relations in case of $n=1$ were proved 
in \cite{LPRS19a}. \qed

The statement of the proposition~\ref{recrel} is a recursion procedure which 
allows  to reduce the set of the Bethe paremetrs $\bar t^{n-1}$ until this set becomes empty.
This yields the presentation of the off-shell $\mathfrak{o}_{2n+1}$-invariant Bethe vector
$\BB(\bar t^{\,0},\ldots, \bar t^{\,n-1})$
as a linear combinations of the products of the matrix entries $T_{i,n}(t^{n-1}_\ell)$, $-n\leq i\leq n-1$
with rational coefficients acting onto 
$\mathfrak{o}_{2n-1}$-invariant Bethe vector
$\BB(\bar t^{\,0},\ldots, \bar t^{\,n-2})$ {(see remark~\ref{embed})}. Then we repeat this procedure to express 
$\mathfrak{o}_{2n-1}$-invariant Bethe vector
$\BB(\bar t^{\,0},\ldots, \bar t^{\,n-2})$
as a linear combinations of the products of the matrix entries $T_{i,n-1}(t^{n-2}_\ell)$, $-n+1\leq i\leq n-2$
acting onto 
$\mathfrak{o}_{2n-3}$-invariant Bethe vector
$\BB(\bar t^{\,0},\ldots, \bar t^{\,n-3})$ and so on. Finally, one may obtain a presentation of the 
off-shell Bethe vector as polynomial of the matrix entries $T_{i,j}(t^{j-1}_\ell)$ with rational coefficients 
acting on the reference vector $\rvac$ with $-j\leq i\leq j-1$ for $1\leq j\leq n$.

Note that  at each step this recurrence procedure is in accordance with embedding of the smaller algebra 
$\Bal_{n-1}$ into the bigger algebra $\Bal_{n}$ described in \cite{JLM18} (see remark~\ref{embed}).

For example, first nontrivial Bethe vectors in $\mathfrak{o}_5$-invariant model 
\begin{equation*}
\BB(t^0;t^1)=\frac{T_{0,2}(t^1)\rvac}{\lambda_{2}(t^1)}+\frac{1}{g(t^1,t^0)}
\frac{T_{1,2}(t^1)T_{0,1}(t^0)\rvac}
{\lambda_{2}(t^1)\lambda_{1}(t^0)}\,,
\end{equation*}

\begin{equation*}
\begin{split}
\BB(t^0;{ \{ } t^1_1,t^1_2 { \} })&=\frac{h(t^1_1,{ t^0})}{h(\bar t^1,\bar t^1)}
\frac{T_{0,2}(t^1_2)T_{1,2}(t^1_1)\rvac}
{\lambda_{2}(t^1_2)\lambda_{2}(t^1_1)}
+\frac{1}{g(t^1_2,{ t^0 })h(\bar t^1,\bar t^1)}
\frac{T_{1,2}(t^1_2)T_{0,2}(t^1_1)\rvac}
{\lambda_{2}(t^1_2)\lambda_{2}(t^1_1)}\\
&+\frac{1}{g(\bar t^1,{ t^0 })h(\bar t^1,\bar t^1)}
\frac{T_{1,2}(t^1_2){ T_{1,2}}(t^1_1)T_{0,1}(t^0 )\rvac}
{\lambda_{2}(t^1_2)\lambda_{2}(t^1_1)\lambda_{1}(t^0_1)}\,,
\end{split}
\end{equation*}
 
\begin{equation*}
\begin{split}
&\BB( { \{ } t^0_1,t^0_2 { \} } ;t^1)=-\frac{T_{-1,2}(t^1)\rvac}{\lambda_{2}(t^1)}+\frac{\fgo(t^0_2,t^0_1)}{g(t^1,t^0_1)}
\frac{T_{0,2}(t^1)T_{0,1}(t^0_1)\rvac}
{\lambda_{2}(t^1)\lambda_{1}(t^0_1)}+\frac{\fgo(t^0_1,t^0_2)}{g(t^1,t^0_2)}
\frac{T_{0,2}(t^1)T_{0,1}(t^0_2)\rvac}
{\lambda_{2}(t^1)\lambda_{1}(t^0_2)}\\
&-\frac{1}{g(t^1,\bar t^0)h(t^0_2,t^0_1)}
\frac{T_{1,2}(t^1)T_{-1,1}(t^0_2)\rvac}
{\lambda_{2}(t^1)\lambda_{1}(t^0_2)}+\frac{1}{g(t^1,\bar t^0)\fgo(t^0_2+c/2,t^0_1)}
\frac{T_{1,2}(t^1)T_{0,1}(t^0_2)T_{0,1}(t^0_1)\rvac}
{\lambda_{2}(t^1)\lambda_{1}(t^0_2)\lambda_{1}(t^0_1)}\,.
\end{split}
\end{equation*}

First nontrivial off-shell Bethe vector in $\mathfrak{o}_7$-invariant model is 
\begin{equation*}
\begin{split}
\BB(t^0;t^1;t^2)&=\frac{T_{0,3}(t^2)\rvac}
{\lambda_{3}(t^2)}+\frac{1}{g(t^1,t^0)}
\frac{T_{1,3}(t^2)T_{0,1}(t^0)\rvac}
{\lambda_{3}(t^2)\lambda_{1}(t^0)}+\frac{1}{g(t^2,t^1)}
\frac{T_{2,3}(t^2)T_{0,2}(t^1)\rvac}
{\lambda_{3}(t^2)\lambda_{2}(t^1)}\\
&+\frac{1}{g(t^2,t^1)g(t^1,t^0)}
\frac{T_{2,3}(t^2)T_{1,2}(t^1)T_{0,1}(t^0)\rvac}
{\lambda_{2}(t^1)\lambda_{1}(t^0)\lambda_{1}(t^0)}\,.
\end{split}
\end{equation*}
To obtain these formulas one has to apply recurrence relations \r{rr1}
and similar relations for  the  $\mathfrak{o}_3$-invariant Bethe vectors
presented in  \cite{LPRS19a}.

Besides recurrence relations given by the proposition~\ref{recrel} corresponding to the last column 
of monodromy matrix one can obtain from the action formulas the recurrence relation with respect 
to the first row of the monodormy matrix. We formulate this relation  as the following
\begin{prop}\label{recrel1}
For $n\geq2$ one has an alternative recurrence relations for the off-shell Bethe vector $\BB(\bar t)$ 
\begin{equation} \label{rr11}
\begin{split}
 &  \BB(\bar t^{\,0},\ldots, 
 \bar t^{\,n-2},\{\bar t^{\,n-1},z_{n-1}\}) = \frac{1}{h(\bar t^{\,n-1} ,z_{n-1})  \lambda_{-n+1}(z)}\\
&\quad\times   \sum_{j=-n+1}^{n} 
\sum_{{\rm part}} 
 \frac{ (-1)^{\delta_{j,n}}}{ h(z_{n-1},\bar t^{\,n-1}_{ \st})}
 \frac  {\sigma_{j}T_{-n ,j}(z)\cdot
 \BB(\bar t_{\st} )} { g(z_{n-2},\bar t^{\,n-2}_{\st})}\ 
     \prod_{s=0}^{n-1}  \alpha_s(\bar t^{\,s}_{\sth}) \gamma_s(\bar t^{\,s}_{\st},\bar t^{\,s}_{\sth}) 
     \prod_{s=1}^{n-1}
    \frac{h(\bar t^{\,s}_{\sth},\bar t^{\,s-1}_{\st})}{g(\bar t^{\,s}_{\st},\bar t^{\,s-1}_{\sth})}  \, ,
    \end{split}
\end{equation}
where sum in \r{rr11} goes over partitions $\{ \bar t^{\,s}_{\st},\bar t^{\,s}_{\sth}\}\vdash \bar t^{\,s}$
with cardinalities  ($s<n-2$) 
\begin{equation*}\label{rr22}
\#\bar t^s_{\sth}=\begin{cases}0,\quad 0\leq s\leq -j-1\,,\\ 1,\quad -j\leq s\leq n-2
\end{cases}
\end{equation*}
for $-n+1\leq j\leq 0$, 
\begin{equation*}\label{rr33}
\#\bar t^s_{\sth}=\begin{cases}2,\quad 0\leq s\leq j-1\,,\\ 1,\quad j\leq s\leq n-2
\end{cases}
\end{equation*}
for $1\leq j\leq n-1$ and  for $s=n-1$
\begin{equation*}\label{rr44}
\#\bar t^{n-1}_{\sth}=\begin{cases}0,\quad -n+1\leq j\leq n-1\,,\\ 1,\quad j=n\,.
\end{cases}
\end{equation*}
\end{prop}

In \r{rr1} and \r{rr11} the sign factor $\sigma_i$ is defined by \r{sign}.

\subsection{Eigenvalue property of the Bethe vectors $\BB(\bar t)$}

Let us demonstrate in this section how the action formulas \r{mainac} reproduce the eigenvalue \r{eigBA}. 
We formulate it as following
\begin{prop}\label{eigenprop}
The action of the transfer matrix \r{trans} onto off-shell Bethe vectors $\BB(\bar t)$ which follows from 
\r{mainac} is 
\begin{equation}\label{eigpr}
\mathcal{T}(z)\cdot \BB(\bar t)=\tau(z;\bar t)\BB(\bar t) +\cdots\,,
\end{equation}
where eigenvalue $\tau(z;\bar t)$ is given by the equality \r{eigBA} and $\ \cdots\ $ stands for the terms
which are vanishing if Bethe equations \r{BE} are satisfied. 
\end{prop}

It is clear that first term in the right hand side of \r{eigpr} 
will correspond to the so called 'wanted' terms in the right hand 
side of the diagonal elements actions $T_{\ell,\ell}(z)$  which corresponds to the partitions 
such that $\bar w_{\st}=\bar t$. It is convenient to consider separately the action 
of $T_{-\ell,-\ell}(z)$ and $T_{\ell,\ell}(z)$ for $\ell=0,1,\ldots,n-1,n$. 
According to \r{card} the cardinalities of the sets $\bar w^s_{\so}$ and $\bar w^s_{\sth}$
for the actions of $T_{-\ell,-\ell}(z)$ are 
\begin{equation*}\label{car1}
\begin{split}
s&=0,\ldots,\ell-1,\quad \#\bar w^s_{\so}=0\quad{\rm and}\quad\#\bar w^s_{\sth}=2\,,\\
s&=\ell,\ldots,n-1,\quad \#\bar w^s_{\so}=1\quad{\rm and}\quad\#\bar w^s_{\sth}=1\,,
\end{split}
\end{equation*}
and for the action of $T_{\ell,\ell}(z)$
\begin{equation*}\label{car2}
\begin{split}
s&=0,\ldots,\ell-1,\quad \#\bar w^s_{\so}=2\quad{\rm and}\quad\#\bar w^s_{\sth}=0\,,\\
s&=\ell,\ldots,n-1,\quad    \#\bar w^s_{\so}=1\quad{\rm and}\quad\#\bar w^s_{\sth}=1\,.
\end{split}
\end{equation*}

In the first case of the action 
$T_{-\ell,-\ell}(z)$ the sets $\bar w^s_{\so}$ and $\bar w^s_{\sth}$ which will correspond to the wanted terms 
are for 
\begin{equation}\label{wan1}
\begin{split}
s&=0,\ldots,\ell-1,\quad \bar w^s_{\so}=\varnothing,\ \ \bar w^s_{\sth}=\{z,z_s\}\,,\\
s&=\ell,\ldots,n-1,\quad \bar w^s_{\so}=\{z\},\  \ \bar w^s_{\sth}=\{z_s\}
\quad{\rm or}\quad\bar w^s_{\so}=\{z_s\},\ \ \bar w^s_{\sth}=\{z\}\,.
\end{split}
\end{equation}
In the second case of the action $T_{\ell,\ell}(z)$ the sets  $\bar w^s_{\so}$ and $\bar w^s_{\sth}$ 
for the wanted terms are 
\begin{equation}\label{wan2}
\begin{split}
s&=0,\ldots,\ell-1,\quad \bar w^s_{\so}=\{z,z_s\},\ \ \bar w^s_{\sth}=\varnothing\,,\\
s&=\ell,\ldots,n-1,\quad \bar w^s_{\so}=\{z\},\ \  \bar w^s_{\sth}=\{z_s\}
\quad{\rm or}\quad\bar w^s_{\so}=\{z_s\},\ \  \bar w^s_{\sth}=\{z\}\,.
\end{split}
\end{equation}
The cases $\ell=0$ in \r{wan1} and in \r{wan2} both correspond to the action of 
$T_{0,0}(z)$ and in this case all sets $\bar w^s_{\so}$ and $\bar w^s_{\sth}$ have 
cardinalities 1. On the other hand the cases $\ell=1,\ldots,n$ in \r{wan1} and in \r{wan2}
correspond to the action of $T_{-\ell,-\ell}(z)$ and $T_{\ell,\ell}(z)$ respectively. 
For the action of  $T_{-\ell,-\ell}(z)$ the term in \r{mainac} which produces the wanted term 
corresponds to the partition $\bar w^s_{\sth}=\{z,z_s\}$ for $s=0,\ldots,\ell-1$ and 
$\bar w^s_{\sth}=\{z\}$, $\bar w^s_{\so}=\{z_s\}$ for $s=\ell,\ldots,n-1$. 
Analogously, for the action of 
$T_{\ell,\ell}(z)$ we have $\bar w^s_{\so}=\{z,z_s\}$ for $s=0,\ldots,\ell-1$ and 
$\bar w^s_{\sth}=\{z\}$, $\bar w^s_{\so}=\{z_s\}$ for $s=\ell,\ldots,n-1$.

Let us calculate the contribution of the action $T_{-\ell,-\ell}(z)$ to the eigenvalue \r{eigBA}. For the 
factors depending on the functional parameters $\lambda_j(z)$ we have 
\begin{equation*}
\begin{split}
&\lambda_{n}(z)\prod_{s=0}^{n-1} \alpha_s(z)\prod_{s=0}^{\ell-1}\alpha_s(z_s)=
\lambda_{n}(z)\prod_{s=0}^{n-1} \frac{\lambda_{s}(z)}{\lambda_{s+1}(z)}
\prod_{s=0}^{\ell-1} \frac{\lambda_{s}(z_s)}{\lambda_{s+1}(z_s)}=\\
&=\frac{\lambda_0(z)\lambda_0(z_0)}{\lambda_{\ell}(z_{\ell-1})}\prod_{s=1}^{\ell-1} \frac{\lambda_{s}(z_s)}{\lambda_{s}(z_{s-1})}
= \frac{1}{\lambda_{\ell}(z_{\ell})}\prod_{s=\ell+1}^{n} \frac{\lambda_{s}(z_{s-1})}{\lambda_{s}(z_s)}=\lambda_{-\ell}(z),
\end{split}
\end{equation*}
where we have used the relations \r{rest} for the functional parameters.
One may further check that  wanted terms of \r{mainac} for the action 
$T_{-\ell,-\ell}(z)$ 
reduces to the function $\lambda_{-\ell}(z)f(\bar t^{\ell-1},z_{\ell-1})f(z_\ell,\bar t_\ell)$ 
restoring part of the eigenvalue \r{eigBA}.
Analogously, one may check that the wanted terms of \r{mainac} for the action of $T_{\ell,\ell}(z)$
produces the term $\lambda_{\ell}(z)f(\bar t^\ell,z)f(z,\bar t^{\ell-1})$ of the eigenvalue \r{eigBA}.

Unfortunately, the  unwanted terms marked by dots in \r{eigpr} cannot be presented in the nice form. One can investigate all these terms case by case and verify that all these terms will be proportional to the 
differences of the left and right hands sides of the equalities \r{BE} and will disappear if the Bethe equations for the  parameters $\bar t$ are satisfied. \qed

\subsection{Reduction to $\mathfrak{gl}_n$-invariant Bethe vectors}

Let us consider formula \r{mainac} for $i=1$, $j=n$ and $\bar t^0=\varnothing$.
For these values of indices $\sigma_1= 1$, $\sigma_{-n}=-1$  and 
\r{card} yields following cardinalities $\#\bar w^s_{\sth}=0$ for 
$s=0,\ldots,n-1$; $\#\bar w^0_{\so}=2$ and $\#\bar w^s_{\so}=1$ for 
$s=1,\ldots,n-1$. Since $\bar w^s_{\sth}=\varnothing$ for all $s=0,\ldots,n-1$ and 
$\bar t^0=\varnothing$, the set $\bar w^0_{\st}=\varnothing$ is empty
and the action formula 
\r{mainac} simplifies to (recall that there is no summation over partition of the set $\bar w^n=\{z,z_n\}$
and $\bar w^n_{\so}=\{z_n\}$, $\bar w^n_{\sth}=\{z\}$, see remark~\ref{reading})
\begin{equation}\label{glnr2}
\begin{split}
    T_{1,n}(z)&\cdot\BB(\varnothing,\bar t^1,\ldots,\bar t^{n-1})=
   - \lambda_{n}(z)\ \kappa\ \frac{h(z,\bar t^{n-1})}{g(z_{n},\bar t^{n-1})}\\
    &\quad\times \sum_{\rm part} \  \BB(\varnothing,\bar w^1_{\st},\ldots,\bar w^{n-1}_{\st})\ 
    \frac{1}{h(\bar w^{n-1}_{\so},z_{n-1})}
\prod_{s=1}^{n-1}\frac{h(\bar w^s_{\st},\bar w^{s-1}_{\so})g(\bar w^s_{\so},\bar w^s_{\st})}
  {h(\bar w^s_{\st},\bar w^{s}_{\so})g(\bar w^s_{\so},\bar w^{s-1}_{\st})}\,.
    \end{split}
\end{equation}
The set $\bar w^0_{\so}$ is equal to $\{z,z_0=z+c/2\}$. Then because of the factor 
$h(\bar w^1_{\st},\bar w^{0}_{\so})$ the sum over partitions of the set $\bar w^1$ reduces
to the single partition $\bar w^1_{\st}=\{\bar t^1,z\}$ and $\bar w^1_{\so}=\{z_1=z-c/2\}$. 
Next the factor  
\begin{equation*}
\frac{h(\bar w^2_{\st},\bar w^1_{\so})}{g(\bar w^2_{\so},\bar w^1_{\st})} =
\frac{h(\bar w^2_{\st},z-c/2)}{g(\bar w^2_{\so},z)g(\bar w^2_{\so},\bar t^1)}
\end{equation*} 
reduces summation over partitions of the set $\bar w^2$ to a single partition 
$\bar w^2_{\st}=\{\bar t^2,z\}$ and $\bar w^2_{\so}=\{z_2=z-3c/2\}$. Continuing 
we find that sum over partitions of all sets $\bar w^s$, $s=1,2,\ldots,n-1$ disappear and 
reduces to a single partition when 
\begin{equation*}
\bar w^s_{\st}=\{\bar t^s,z\}\quad{\rm and}\quad \bar w^s_{\so}=\{z_s\}\quad s=1,2,\ldots,n-1\,.
\end{equation*}

For these sets and $\bar w^0_{\so}=\{z,z_0=z+c/2\}$, $\bar w^0_{\st}=\varnothing$ 
the product in \r{glnr2} is equal to 
\begin{equation*}
\frac{1}{h(\bar w^{n-1}_{\so},z_{n-1})}
\prod_{s=1}^{n-1}\frac{h(\bar w^s_{\st},\bar w^{s-1}_{\so})g(\bar w^s_{\so},\bar w^s_{\st})}
  {h(\bar w^s_{\st},\bar w^{s}_{\so})g(\bar w^s_{\so},\bar w^{s-1}_{\st})}
=-\frac{1}{\kappa}\ h(\bar t^1,z) g(z_n,\bar t^{n-1})\,.
\end{equation*}

Summarizing we can write the action \r{glnr2} in the form 
\begin{equation}\label{glnr1}
    T_{1,n}(z)\cdot\BB(\varnothing,\bar t^1,\ldots,\bar t^{n-1})=
     \lambda_{n}(z)h(\bar t^1,z)h(z,\bar t^{n-1})
    \BB(\varnothing, \bet^1,\ldots,\bet^{n-1})\,,
\end{equation}
where $\bet^s=\{\bar t^s,z\}$, $s=1,\ldots,n-1$. This action coincides with the action 
given by lemma~4.2 of the paper \cite{HutLPRS20}. Indeed, if we renormalize 
the Bethe vectors of this paper
\begin{equation}\label{renorm}
    \tilde\BB(\bar t^1,\ldots,\bar t^{n-1})=\frac{\prod_{s=1}^{n-1} h(\bar t^s,\bar t^{s})}
    {\prod_{s=2}^{n-1} h(\bar t^s,\bar t^{s-1})} \BB(\varnothing,\bar t^1,\ldots,\bar t^{n-1})
\end{equation}
we obtain from \r{glnr1}
\begin{equation}\label{gln-act}
    T_{1,n}(z)\cdot\tilde\BB(\bar t^1,\ldots,\bar t^{n-1})=\lambda_{n}(z)\ 
    \tilde\BB(\{\bar t^1,z\},\ldots,\{\bar t^{n-1},z\}),
\end{equation}
which  literally coincide with the action of monodromy matrix entry $T_{1,n}(z)$ 
onto off-shell Bethe vectors $\tilde\BB(\bar t)$  in $\mathfrak{gl}_n$-invariant integrable models. 

Let us note that formulas \r{ZM111} at $\bar t^0=\varnothing$ for 
$i=1,\ldots,n-1$ 
yields the equality (4.3) of the paper \cite{HutLPRS20} 
\begin{equation}\label{glnr3}
\begin{split}
& T_{i+1,i}\cdot \tilde\BB(\bar t^1,\ldots,\bar t^{n-1})=\sum_{\ell=1}^{r_i} 
   \tilde\BB(\bar t^1,\ldots,\bar t^{\,i-1},\bar t^{\,i}_\ell; \bar t^{\,i+1},\ldots,\bar t^{\,n-1})\\
&\quad\times \sk{\kk_{i+1}\ \frac{\lambda_{i}(t^i_\ell)}{\lambda_{i+1}(t^i_\ell)}   
 \frac{f(\bar t_\ell^{\,i},t^{i}_\ell)}{f(\bar t^{\,i+1},t^i_\ell)}
 -  \kk_{i}\  \frac{f(t^{i}_\ell,\bar t_\ell^{\,i})}  {f(t^i_\ell,\bar t^{\,i-1})} }
\end{split}
\end{equation}
for renormalized  vectors $\tilde\BB(\bar t)$ \r{renorm}. 

Since the action formulas \r{gln-act} and \r{glnr3} coincide with the statements of the lemma 4.2 of the 
paper  \cite{HutLPRS20} they can be used to restore the action of entries $T_{i,j}(z)$ onto 
Bethe vectors $\tilde\BB(\bar t^1,\ldots,\bar t^{n-1})$ in the framework of zero-modes method. 
So far we have demonstrated that the  action formulas \r{mainac} of $T_{i,j}(z)$ for 
$1\leq i,j\leq n$ onto off-shell Bethe vectors  
 $\BB(\varnothing,\bar t^1,\ldots,\bar t^{n-1})$ with empty set  
 $\bar t^0=\varnothing$
 leads to the action formulas in 
$\mathfrak{gl}_n$-invariant models which was calculated in \cite{HutLPRS20}.

Let us also check that recurrence relations given by the proposition~\ref{recrel} yield the 
corresponding relations found  in \cite{HLPRS17b}. Formula \r{rr1} in case of $\bar t^0=\varnothing$
becomes
\begin{equation*}\label{ren1}
\begin{split}
\BB(\varnothing,\bar t^1,\ldots,\bar t^{n-2},\{\bar t^{n-1},z\})&=\sum_{i=1}^{n-1}\sum_{{\rm part}}
\frac{T_{i,n}(z)\cdot \BB(\varnothing,\bar t^1,\ldots,\bar t^{i-1},\bar t^i_{\st},\ldots,\bar t^{n-2}_{\st},\bar t^{n-1})}
{\lambda_n(z)h(z,\bar t^{n-1})h(\bar t^{n-1},z)g(z,\bar t^{n-2}_{\st})}\\
&\quad\times 
\prod_{s=i}^{n-2}\frac{g(\bar t^s_{\so},\bar t^s_{\st})}{h(\bar t^s_{\st},\bar t^s_{\so})}\ 
\frac{h(\bar t^{n-1},\bar t^{n-2}_{\so})}{g(\bar t^i_{\so},\bar t^{i-1})}\ 
\prod_{s=i+1}^{n-2}\frac{h(\bar t^s_{\st},\bar t^{s-1}_{\so})}{g(\bar t^s_{\so},\bar t^{s-1}_{\st})}\,.
\end{split}
\end{equation*}
The sum over $i$ in \r{rr1} reduces to the interval $1\leq i\leq n-1$ and 
we have to take into account \r{rr2} which states that $\bar t^s_{\so}=\varnothing$ for $1\leq s\leq i-1$ and 
$\bar t^{n-1}_{\so}=\varnothing$ according to \r{rr4}. 
This recurrence relation for the renormalized Bethe vector \r{renorm} 
\begin{equation*}\label{ren2}
\begin{split}
\tilde\BB(\bar t^1,\ldots,\bar t^{n-2},\{\bar t^{n-1},z\})&=\sum_{i=1}^{n-1}
\frac{T_{i,n}(z)}{\lambda_n(z)}
\sum_{{\rm part}}
 \tilde\BB(\bar t^1,\ldots,\bar t^{i-1},\bar t^i_{\st},\ldots,\bar t^{n-2}_{\st},\bar t^{n-1})\\
&\quad\times 
\frac{g(z,\bar t^{n-2}_{\so})}{f(z,\bar t^{n-2})}\ 
\frac{\prod_{s=i}^{n-2}f(\bar t^s_{\so},\bar t^s_{\st}) \prod_{s=i+1}^{n-2} 
g(\bar t^s_{\so},\bar t^{s-1}_{\so}) }
{\prod_{s=i}^{n-2} f(\bar t^s_{\so},\bar t^{s-1})}
\end{split}
\end{equation*}
coincides identically with equation (4.4) of the paper \cite{HLPRS17b} in case $m=0$.

\subsection{Action formulas for $\mathfrak{o}_3$-invariant models}

In \cite{LPRS19} we have calculated the action formulas 
for $\mathfrak{o}_3$-integrable model. Let us verify that general formula \r{mainac}
reduces to these action formulas at $n=1$. Cardinalities of the sets $\bar w^0_{\so}$ and 
$\bar w^0_{\sth}$ according to \r{card} will be in this case $(-1\leq i,j\leq 1)$
\begin{equation*}
\begin{split}
\#\bar w^0_{\so}&=\Theta(i)+\Theta(i-1)=i+1\,,\\
\#\bar w^0_{\sth}&=\Theta(-j)+\Theta(-j-1)=1-j\,.
\end{split}
\end{equation*}
Equality  \r{mainac} reduces to 
\begin{equation*}
T_{i,j}(z)\cdot \BB(\bar t^0)=(-1)^{\delta_{i,-1}+\delta_{j,-1}}\frac{\lambda_{1}(z)}{2}
\sum_{\rm part}\  \BB(\bar w^{0}_{\st})\ 
\frac{\alpha_0(\bar w^0_{\sth}) \fgo(\bar w^0_{\so},\bar w^0_{\st}) \fgo(\bar w^0_{\st},\bar w^0_{\sth})
\fgo(\bar w^0_{\so},\bar w^0_{\sth}) }
{h(z,\bar w^0_{\sth})h(\bar w^0_{\so},z+c/2)}
\end{equation*}
which coincides exactly with the action calculated in the paper \cite{LPRS19}
and 
$\bar w^0=\{\bar t^0,z,z+c/2\}$. Here we used the fact that  for $-1\leq i\leq1$,
 $\sigma_i(-1)^{i}=\sigma_{-i}$ and $\sigma_i=(-1)^{\delta_{i,1}+1}$.

\section{Bethe vectors and  algebra $\DYBn$}\label{BVvsC}

As we already mentioned above in order to prove the propositions~\ref{ZMp} and \ref{achimo}
we identify the $RTT$ algebra $\Bal$ with the Borel subalgebra in the 
Yangian double $\DYBn$. This infinite dimensional
algebra  is generated by two $T$-operators $T^\pm(u)$ which satisfy the  
commutation relations 
\begin{equation}\label{rtt-dy}
  R(u,v) \left( T^\mu(u)\otimes\mathbf{I} \right) \left( \mathbf{I}\otimes T^\nu(v) \right) =
  \left( \mathbf{I}\otimes T^\nu(v) \right) \left( T^\mu(u)\otimes\mathbf{I} \right) R(u,v) 
\end{equation}
with the same $R$-matrix \r{Rmat} for $\mu,\nu=\pm$. We set the corresponding central 
elements  given by the equality \r{qu-rel} for $T$-operators $T^\pm(u)$ equal to 1.
We assume following series expansion of the generating series $T^\pm_{i,j}(u)$ 
\begin{equation}\label{dep-yd}
T^\pm_{i,j}(u)=\twis_i\delta_{ij}\pm\sum_{\genfrac{}{}{0pt}{2}{\ell\geq0}{\ell<0}} T_{i,j}[\ell](u/c)^{-\ell-1}\,.
\end{equation}
This expansion allows to identify monodromy matrix $T(u)$ of $\mathfrak{o}_{2n+1}$-invariant
model with $T$-operator $T^+(u)$.

Algebra $\DYBn$ has also so called 'current' realization \cite{D88} described in details in 
\cite{JLY18,LP20}. The link between $RTT$ and 'current' realizations is established through 
{\it Gauss coordinates} which can be defined as follows
\begin{equation}\label{Gauss}
T^\pm_{i,j}(u)=\sum_{{\rm max}(i,j)\leq s\leq n} \FF^\pm_{s,i}(u)k^\pm_s(u)\EE^\pm_{j,s}(u)\,,
\end{equation}
where we assume that $\FF^\pm_{i,j}(u)=\EE^\pm_{j,i}(u)=0$ for $i<j$ and $\FF^\pm_{i,i}(u)=\EE^\pm_{i,i}(u)=1$ for 
$i=-n,\ldots,n$. It was shown in \cite{JLY18,LP20} that 
the set of the Gauss coordinates
\begin{equation*}\label{choice1}
\FF^\pm_{i+1,i}(u),\quad \EE^\pm_{i,i+1}(u),\quad 1\leq i\leq n-1\,,\quad
k^\pm_j(u),\quad 1\leq j\leq n
\end{equation*}
can be chosen 
as the algebraically independent set of the generators of the Yangian double $\DYBn$. 
Commutation relations of this algebra contains also currents $k^\pm_0(u)$. The modes 
of these currents $k_0[\ell]$, $\ell\in\ZZ$ can be expressed through modes  
$k_s[\ell]$, $\ell\in\ZZ$, $s=1,\ldots,n$ of the algebraically independent currents 
by the relations
\begin{equation*}\label{center}
k^\pm_{0}(u+c/2)k^\pm_{0}(u)=
\prod_{s=1}^{n}\frac{k^\pm_{s}(u-c(s-3/2))}{k^\pm_{s}(u-c(s-1/2))}\,.
\end{equation*}

$RTT$ commutation relations \r{rtt-dy} for the Yangian double $\DYBn$ can be presented 
in terms of the formal generating series or  {\it currents} 
\begin{equation}\label{DF-iso1}
\begin{split}
F_i({u})&=\FF^{+}_{i+1,i}({u})-\FF^{-}_{i+1,i}({u})=\sum_{\ell\in\ZZ}F_i[\ell]u^{-\ell-1}\,,\\
E_i({u})&=\EE^{+}_{i,i+1}({u})-\EE^{-}_{i,i+1}({u})=\sum_{\ell\in\ZZ}E_i[\ell]u^{-\ell-1}
\end{split}
\end{equation}
for $0\leq i\leq n-1$. We will write explicitly only those 
commutation relations in terms of the currents 
which will be relevant for the calculations (see, for example,  \r{tFiFi}, \r{tkiF} and \r{ZM11} below). 
One can find the whole set of the relations between currents in  \cite{JLY18,LP20}.

$RTT$ realization of the Yangian double and its 'current' realization correspond to 
the different choices of Borel subalgebras. Following the ideas of the paper \cite{EKhP07} 
one can define projections onto intersections of these different type Borel subalgebras. 
As it was shown in several papers (see, for example, \cite{HutLPRS17a} for the super-symmetric
Yangian double $\mathcal{D}Y(\mathfrak{gl}(m|n))$ and references therein) these 
projections being applied to the products of the currents $F_i(u)$ \r{DF-iso1} 
can be identified with off-shell Bethe vectors in the corresponding $\mathfrak{g}$-invariant 
integrable models.

In \cite{LP20} one can find a formal definition of the projections onto intersections of the different type Borel subalgebras 
in the Yangian doubles. Very often one can use 
following approach to calculate these projections of the product of the currents $F_i(u)$. 
One should replace each current by the difference of the Gauss coordinates according to 
the Ding-Frenkel formulas \r{DF-iso1}, expand all brackets and use commutation relations 
between Gauss coordinates which follow from \r{rtt-dy} to order all monomials in such a way 
that all 'positive' Gauss coordinates $\FF^+_{j,i}(u)$, $i<j$ are on the right of all 'negative' 
Gauss coordinates  $\FF^-_{j,i}(u)$ in each monomial. Although we start from the Gauss coordinates $\FF^\pm_{j,i}(u)$
for $j=i+1$ the higher Gauss coordinates for $j>i+1$ will appear during the process of the ordering.
Then application of the projection $\Pfp$ amounts to  delete all 
ordered monomials composed from the Gauss coordinates which have at least one 'negative' 
Gauss coordinate $\FF^-_{j,i}(u)$ on the left. Analogously, application of the projection $\Pfm$ amounts 
to delete all monomials which have at least one 'positive' Gauss coordinate  $\FF^+_{j,i}(u)$
on the right. Analogously, one can define the applications of the projections $\Pepm$ to the 
products of the currents $E_i(u)$.

Let $\bar t$ be a set of the generic parameters described in   \r{sets}.
For any scalar function $x(t,t')$ of two variables and a set 
$\bar t^{\,s}=\{t^s_1,\ldots,t^s_{r_s}\}$, $r_s>1$  
we introduce the 'triangular' products of these functions 
\begin{equation}\label{t-p}
\Defun{x}( \bar t^{s}  )=\prod_{\ell>\ell'}x(t^s_\ell,t^s_{\ell'})\,.
\end{equation}
For any $s=0,\ldots,n-1$ and the set $\bar t^{\,s}$ of the Bethe parameters we define the 
normalized ordered products of the currents 
\begin{equation*}\label{sym-pr}
\F_{s}(\bar t^{\,s})=\Defun{f_s}(\bar t^{\,s}) F_{s}(t^s_{r_s})\cdots F_{s}(t^s_1)\quad\mbox{for}\quad
s=0,1,\ldots,n-1\,,
\end{equation*}
where $f_s$ are defined by \eqref{fun-go}.
By definition we set $\F_{s}(\varnothing)=1$ for all $s$. 
It is clear from the commutation relation
\begin{equation}\label{tFiFi}
f_s(t,t')\ F_s(t)F_s(t')=  f_s(t',t)\  F_s(t')F_s(t),\quad 0\leq s\leq n-1
\end{equation}
 that these ordered products are 
symmetric with respect to permutations of the elements in any of the set $\bar t^{\,s}$.

Let $\bar t$ be a set of generic parameters described by \r{sets}.
Let us define the ordered products of the currents
\begin{equation}\label{nk1}
    \BF^0(\bar t)=\F_{n-1}(\bar t^{\,n-1})\F_{n-2}(\bar t^{\,n-2})\cdots 
    \F_{1}(\bar t^{\,1})\F_{0}(\bar t^{\,0})
\end{equation}
and 
\begin{equation}\label{dres}
\BF(\bar t)=\prod_{s=1}^{n-1}\frac{1}{g(\bar t^{\,s},\bar t^{\,s-1})}
\frac{1}{h(\bar t^{\,s},\bar t^{\,s})}\ \BF^0(\bar t).
\end{equation}
\begin{Def}\label{BVdef}
The $\mathfrak{o}_{2n+1}$-invariant off-shell Bethe vector $\BB(\bar t)$ is defined 
by the action of the projection $\Pfp\sk{\BF(\bar t)}$ onto reference vector $\rvac$ \r{vac}
\begin{equation}\label{beth1}
    \BB(\bar t)=\Pfp\sk{\BF(\bar t)}\rvac.
\end{equation}
\end{Def}

Products of the currents  \r{nk1} and \r{dres} can be written in the ordered form \cite{LP20}
\begin{equation}\label{Bn-co}
    \BF^0(\bar t)=\sum_{{\rm part}}\fgo(\bar t^{\,0}_{\so},\bar t^{\,0}_{\st}) 
    \prod_{s=1}^{n-1} f(\bar t^{\,s}_{\so},\bar t^{\,s}_{\st})
f(\bar t^{\,s}_{\st},\bar t^{\,s-1}_{\so})
\Pfm\sk{\BF^0(\bar t_{\so})}\cdot \Pfp\sk{\BF^0(\bar t_{\st})}
\end{equation}
and
\begin{equation}\label{Bnor-co}
\begin{split}
    \BF(\bar t)&=\sum_{{\rm part}}\fgo(\bar t^{\,0}_{\so},\bar t^{\,0}_{\st}) 
    \prod_{s=1}^{n-1}\frac{ g(\bar t^{\,s}_{\so},\bar t^{\,s}_{\st})
h(\bar t^{\,s}_{\st},\bar t^{\,s-1}_{\so})}
{ h(\bar t^{\,s}_{\st},\bar t^{\,s}_{\so})
g(\bar t^{\,s}_{\so},\bar t^{\,s-1}_{\st})}
\Pfm\sk{\BF(\bar t_{\so})}\cdot \Pfp\sk{\BF(\bar t_{\st})}\\
&=\sum_{{\rm part}}
    \prod_{s=0}^{n-1}
    \gfu_s(\bar t^{\,s}_{\so},\bar t^{\,s}_{\st})
  \prod_{s=1}^{n-1}\frac{ h(\bar t^{\,s}_{\st},\bar t^{\,s-1}_{\so})}
{ g(\bar t^{\,s}_{\so},\bar t^{\,s-1}_{\st})}
\Pfm\sk{\BF(\bar t_{\so})}\cdot \Pfp\sk{\BF(\bar t_{\st})},
\end{split}
\end{equation}
where functions $\gfu_s(u,v)$ are defined by \r{t-fun}
and 
summation goes over all possible partitions of the sets $\bar t^{\,s}$ onto nonintersecting 
subsets $\bar t^{\,s}_{\so}$ and $\bar t^{\,s}_{\st}$ such that 
$\{\bar t^{\,s}_{\so},\bar t^{\,s}_{\st}\}\vdash\bar t^{\,s}$ and 
$\#\bar t^{\,s}_{\so}+\#\bar t^{\,s}_{\st}=\#\bar t^{\,s}$.
Cardinalities of the sets  $\bar t^{\,s}_{\so}$ and $\bar t^{\,s}_{\st}$ can be equal to zero. 
Notations $\bar t_{\so}$ and $\bar t_{\st}$ means the collections of the subsets 
$\bar t^{\,s}_{\so}$ and $\bar t^{\,s}_{\st}$
\begin{equation*}
    \bar t_{\so}=\{\bar t^{\,0}_{\so},\bar t^{\,1}_{\so},\ldots,\bar t^{\,n-1}_{\so}\}\quad
    \mbox{and}\quad 
     \bar t_{\st}=\{\bar t^{\,0}_{\st},\bar t^{\,1}_{\st},\ldots,\bar t^{\,n-1}_{\st}\}.
\end{equation*}
We assume that $\BF^0(\varnothing)\equiv\BF(\varnothing)\equiv 1$ and $\Pfpm(1)=1$. 

\subsection{Proof of the proposition~\ref{ZMp}}

Equations \r{dep-yd} and   \r{Gauss} yield the following series expansion
\begin{equation*}\label{ZM8}
k^\pm_j(u)=\kk_j\pm\sum_{\ell\geq 0\atop \ell<0} k_j[\ell](u/c)^{-\ell-1} .
\end{equation*}
Because of this  expansion 
the zero-modes of  $T^+_{j+1,j}(u)$ for $0\leq j\leq n-1$ is 
\begin{equation*}\label{ZM9}
T^+_{j+1,j}=\kk_{j+1}E_{j}[0],
\end{equation*}
where $E_j[0]$ are the zero-modes of the  currents $E_j(z)$ for $j=0,\ldots,n-1$ corresponding to the simple roots
of the algebra $\mathfrak{o}_{2n+1}$. 

To calculate the action of  the operator $E_j[0]$ onto off-shell Bethe vector we use a special
presentation for the ordered product of the currents $\BF(\bar t)$ \r{dres} which follows from 
\r{Bnor-co}
\begin{equation}\label{ZM10} 
\begin{split}
&\Pfp\sk{\BF(\bar t)}=\BF(\bar t)+\sum_{\ell=1}^{r_0} 
\frac{\fgo(t_\ell^0,\bar t^{\,0}_\ell)f(\bar t^{\,1},t^0_\ell)}
{g(\bar t^{\,1},t^0_\ell)}\FF^-_{1,0}(t^0_\ell)
\BF(\bar t^{\,0}_\ell;\bar t^{\,1};\ldots;\bar t^{\,n-1})+\\
&+\sum_{s=1}^{n-1}\sum_{\ell=1}^{r_s} 
\frac{f(t_\ell^s,\bar t^{\,s}_\ell)f(\bar t^{\,s+1},t^s_\ell)}
{g(\bar t^{\,s+1},t^s_\ell)g(t^s_\ell,\bar t^{\,s-1})h(\bar t^{\,s},t^s_\ell)h(t^s_\ell,\bar t^{\,s})}
\FF^-_{s+1,s}(t^s_\ell)
\BF(\bar t^{\,0};\ldots;\bar t^{\,s}_\ell;\ldots;\bar t^{\,n-1})
+ \cdots {\blu \,,}
\end{split}
\end{equation}
where $\cdots$ stands for terms which are  annihilated by the projection $\Pfp$
after the adjoint action of the zero mode $E_j[0]$.

Using  the commutation relations 
\begin{equation}\label{tkiF}
\begin{split}
k^{\pm}_{0}(u)F_{0}(v)k^{\pm}_{0}(u)^{-1}&=f(u,v)f(v,u+c/2)F_{0}(v),\\
k^{\pm}_i(u) F_i(v) k^{\pm}_i(u)^{-1}&= f(v,u)\ F_i(v),\quad 1\leq i\leq { n-1},\\
k^{\pm}_{i+1}(u)F_i(v)k^{\pm}_{i+1}(u)^{-1}&= f(u,v)\ F_i(v),\quad 0\leq i\leq { n-1},\\
k^{\pm}_i(u) F_j(v) k^{\pm}_i(u)^{-1}&= F_j(v),\quad j\neq i,i+1,\quad  0\leq j\leq n-1, \quad 0\leq i\leq n 
 \end{split}
\end{equation} 
and formulas 
\begin{equation*}\label{ZM11}
\begin{split}
E_{j}[0]F_{l}(v) &=F_{l}(v) E_{j}[0] + \delta_{jl}
\sk{k^+_j(v)k^+_{j+1}(v)^{-1}- k^-_j(v)k^-_{j+1}(v)^{-1}},\\[2mm]
E_{j}[0]\FF^-_{l+1,l}(v) &=\FF^-_{l+1,l}(v) E_{j}[0] + \delta_{jl}
\sk{k^-_j(v)k^-_{j+1}(v)^{-1}- \kk_j\kk_{j+1}^{-1}},
\end{split}
\end{equation*}
where $j,l=0,\ldots,n-1$ 
we may calculate for $i=0,\ldots,n-1$
\begin{equation}\label{ZM111}
\begin{split}
    &T_{i+1,i}\cdot \BB(\bar t)=\sum_{\ell=1}^{r_i} 
    \BB(\bar t^{\,0};\ldots;\bar t^{\,i-1};\bar t^{\,i}_\ell;\bar t^{\,i+1};\ldots;\bar t^{\,n-1})\\
&\quad\times 
    \sk{\kk_{i+1}\ \frac{\lambda_{i}(t^i_\ell)}{\lambda_{i+1}(t^i_\ell)}-
    \kk_{i}\  \frac{f_i(t^{i}_\ell,\bar t_\ell^{\,i})f(\bar t^{\,i+1},t^i_\ell)} 
    {f_i(\bar t_\ell^{\,i},t^{i}_\ell)f(t^i_\ell,\bar t^{\,i-1})} }
  \frac{f_i(\bar t_\ell^{\,i},t^{i}_\ell)}
{g(\bar t^{\,i+1},t^i_\ell)}
 \frac{h(t^i_\ell,\bar t^{\,i-1})}
{ h(t^i_\ell,\bar t^{\,i})h(\bar t^{\,i},t^{i}_\ell)}.
\end{split}
\end{equation}
In \r{ZM10} and \r{ZM111} we set $\bar t^{-1}=\bar t^{n}=\varnothing$ and the products of the rational functions depending 
on these sets are equal to 1. 

If $\kk_j=1$ and the Bethe parameters $\bar t$  satisfy the Bethe equations \r{BE}
the on-shell Bethe vectors become highest weight vectors for the algebra $\mathfrak{o}_{2n+1}$.  
Performing calculations of the zero modes actions \r{ZM111}  we omit the terms 
which are annihilated by the action of the projection $\Pfp$.  
In particular, the terms containing the ratio of $ {k^-_j(v)}/{k^-_{j+1}(v)}$ do not contribute to these actions.

To prove proposition~\ref{ZMp} it is sufficient to use the action \r{ZM111} of the 
$\mathfrak{o}_{2n+1}$ simple roots monodromy 
matrix elements zero modes and the commutation relations \r{ZM777}. 
 \qed

\subsection{Proof of the proposition~\ref{achimo}}

According to Gauss decomposition \r{Gauss}
the monodromy matrix element $T^+_{-n,n}(z)$ is
\begin{equation}\label{extT1}
    T^+_{-n,n}(z)=\FF^+_{n,-n}(z)k^+_{n}(z)\,.
\end{equation}
We will calculate the action of this  element 
$T^+_{-n,n}(z)$ onto ordered product of the currents  $\Pfp\sk{\BF^0(\bar t)}$. 
We need following 
\begin{lemma}\label{T1NFm}
In the subalgebra $\Bal$ and for $-n\leq i<j\leq n$ we have the equality 
\begin{equation}\label{T1Nac}
\Pfp\sk{T^+_{-n,n}(v)\FF^-_{j,i}(u)}=0\,.
\end{equation}
\end{lemma}

Let us consider the equality \r{rtt-dy} 
for the values $\mu=-$ and $\nu=+$ and multiply it from the right by $R(v,u)$. Using 
unitarity condition \r{Runi} we obtain 
\begin{equation*}\label{dy1}
\sk{1-\frac{c^2}{(u-v)^2}} \left( \mathbf{I}\otimes T^+(v) \right) \left( T^-(u)\otimes\mathbf{I} \right)=
   R(u,v) \left( T^-(u)\otimes\mathbf{I} \right) \left( \mathbf{I}\otimes T^+(v) \right)  R(v,u) .
\end{equation*}
If we consider the element $(i,j;-n,n)$ with $-n\leq i<j\leq n$ in this matrix equation we obtain 
\begin{equation}\label{ttr7}
\begin{split}
&\left( 1 - \frac{c^2}{(v-u)^2}\right) T^+_{-n,n}(v) T^-_{i,j}(u)   =  T^-_{i,j}(u) T^+_{-n,n}(v)  
+ \frac{c}{u-v}   T^-_{-n,j}(u) T^+_{i,n}(v) +\\ 
&\qquad+\frac{c}{v-u} \left( T^-_{i,n}(u) T^+_{-n,j}(v)  + \frac{c}{u-v}   T^-_{-n,n}(u) T^+_{i,j}(v)\right).
\end{split}
\end{equation}
Multiplying both sides of the equality \r{ttr7}  on the right by $k^-_{j}(u)^{-1}$ and ordering all terms 
according to the circular ordering in the   Yangian double $\DYBn$ described in 
\cite{LP20} we obtain the statement of the  lemma. 
All terms in RHS of \r{ttr7} start with some Gauss coordinates $\FF^-_{j,i}(t)$, 
so they annihilated by the action of projection $\Pfp$.
\qed

The total currents $F_i(u)$  \r{DF-iso1} are defined for the values $i=0,\ldots,n-1$. 
It was shown in \cite{LP20} using results of the paper \cite{LPRS19} that the same differences 
of the Gauss coordinates defines the currents for the values $i=-n,\ldots,-1$ 
\begin{equation*}\label{co5}
F_i(u)=-F_{-i-1}(u+c(i+3/2))\,.
\end{equation*}
It was proved further in \cite{LP20} that the Gauss coordinates of $T$-operators $T^\pm(u)$ 
are related to the currents $F_i(u)$, for $i=-n,\ldots,n-1$ 
\begin{equation}\label{GC8F}
\FF^+_{j,i}(v)=\Pfp\sk{F_{i}(v)\cdot F_{i+1}(v)\cdots F_{j-2}(v)\cdot F_{j-1}(v)}\,,
\end{equation}
 \begin{equation}\label{8Fm}
\tFF^-_{j,i}(v)=\Pfm\sk{F_{i}(v)\cdot F_{i+1}(v)\cdots F_{j-2}(v)\cdot F_{j-1}(v)}\,,
\end{equation}
where $-n\leq i<j\leq n$ and 
\begin{equation*}\label{tFF}
\tFF^\pm_{j,i}(u) =\sum_{\ell=0}^{j-i-1}(-)^{\ell+1}
\sum_{j>i_\ell>\cdots>i_1>i} \FF^\pm_{i_1,i}(u)  \FF^\pm_{i_2,i_1}(u)\cdots \FF^\pm_{i_\ell,i_{\ell-1}}(u)  
\FF^\pm_{j,i_\ell}(u)\,.
\end{equation*}

Let us introduce the set 
$\{z^0,z^1,\ldots,z^{n-1}\}$  
of the generic complex parameters such that $z^s\not=z^{s'}$  if $s\not=s'$ for  all $s,s'=0,1,\ldots, n-1$ and 
\begin{equation}\label{zsh}
    \tz^i=z^i-c\sk{i-\frac{1}{2}}\,.
\end{equation}
Then using \r{GC8F} we may express monodromy element entry \r{extT1} through the currents 
as follows
\begin{equation*}\label{extT}
    T^+_{-n,n}(z)
    =(-1)^n\Pfp\left.\sk{F_{n-1}(\tz^{n-1})\cdots F_{0}(\tz^0)\cdot
    F_{0}(z^0)\cdots F_{n-1}(z^{n-1})}k^+_{n}(z^{n-1})\right|_{z^i=z}\,,
\end{equation*}
where parameters $\tz^i$ are defined by \r{zsh} for $i=0,\ldots,n-1$. 

Using the statement of the lemma~\ref{T1NFm}, the equality  \r{Bn-co} 
and the fact that 'negative' projection $\Pfm\sk{\BF^0(\bar t_{\so})}$ in this equality
can be expressed in terms of the 'negative' Gauss coordinates $\FF^-_{j,i}(u)$ due to \r{8Fm} 
we can calculate the action of matrix entries $T^+_{-n,n}(z)$ onto 
ordered product of the currents $\Pfp\sk{\BF^0(\bar t)}$
 as follows
\begin{equation}\label{ex-ac}
    T^+_{-n,n}(z)\cdot \Pfp\sk{\BF^0(\bar t)}=\Pfp\sk{ T^+_{-n,n}(z)\cdot \BF^0(\bar t)}\,.
\end{equation}

Let us define an element $T^0_{-n,n}(z^0,\ldots,z^{n-1})$
\begin{equation*}\label{Texc}
    T^0_{-n,n}(z^0,\ldots,z^{n-1})=(-1)^n F_{n-1}(\tz^{n-1})\cdots F_{0}(\tz^0)\cdot
    F_{0}(z^0)\cdots F_{n-1}(z^{n-1})k^+_{n}(z^{n-1}).
\end{equation*}
Then the action \r{ex-ac} can be written as 
\begin{equation*}\label{ex-ac1}
    T^+_{-n,n}(z)\cdot \Pfp\sk{\BF^0(\bar t)}=
    \left.\Pfp\sk{ T^0_{-n,n}(z^0,\ldots,z^{n-1})\cdot \BF^0(\bar t)}\right|_{z^i=z}\,.
\end{equation*}
Using the commutation relations between currents \r{tFiFi}, \r{tkiF}
and 
\begin{equation*}\label{tFiFii}
(t-t'-c)\ F_i(t)F_{i+1}(t')= (t-t')\ F_{i+1}(t')F_i(t),\quad 0\leq i\leq n-2
\end{equation*}
we can reorder 
the currents in the element $T^0_{-n,n}(z^0,\ldots,z^{n-1})\cdot \BF^0(\bar t)$
as follows
\begin{equation}\label{reord}
    T^0_{-n,n}(z^0,\ldots,z^{n-1})\cdot \BF^0(\bar t)=\frac{(-1)^n}{2}
    \frac{\BF^0(\bar w)  }{\prod_{s=1}^{n-1} f(z^{s},z^{s-1})f(\bar t^{\,s},z^{s-1})f(z^{s-1},\bar t^{\,s-1})}
    k^+_{n}(z^{n-1}) +\cdots \,,
\end{equation}
 where $\cdots$ stands for the terms which have higher zeros when 
$z^{\ell}=z^{\ell'}=z$ and do not contribute into \r{dr1}.
The collection of sets $\bar w$ in \r{reord} is
\begin{equation*}\label{wsets}
\bar w=\{\bar w^{ 0},\ldots,\bar w^{ n-1}\}\quad \mbox{and}\quad 
\bar w^i=\{\bar t^{\,i},z^i,\tz^i\}.
\end{equation*}
To obtain \r{reord} we used  a trivial identity 
\begin{equation*}
    f(u-c,v)f(v,u)=1
\end{equation*}
and the fact that we may replace any $z^\ell$ by any $z^{\ell'}$ since at the end we will set 
$z^{\ell}=z^{\ell'}=z$. 

It is obvious that we cannot set $z^\ell=z$ in \r{reord} since $f(z,z)^{-1}=0$. 
But the action of $T^0_{-n,n}(z^0,\ldots,z^{n-1})$ onto renormalized ordered product of currents 
$\BF(\bar t)$ \r{dres} 
is non-singular and yields the action of 
$T^+_{-n,n}(z)$ 
onto off-shell Bethe vector $\BB(\bar t)$ \r{beth1} 
\begin{equation}\label{dr1}
\begin{split}
&T^+_{-n,n}(z)\cdot \BB(\bar t) = T^+_{-n,n}(z)\cdot \Pfp\sk{\BF(\bar t)} \rvac=
\Pfp\sk{T^+_{-n,n}(z)\cdot \BF(\bar t)} \rvac=\\
&\quad =    \left.\Pfp\sk{  T^0_{-n,n}(z^0,\ldots,z^{n-1})\cdot \BF(\bar t)}\right|_{z^i=z}\rvac   =
  -\kappa\  \frac{g(z_1,\bar t^{\,0}) h(z,\bar t^{\,n-1})}{
    \textstyle{h(z,\bar t^{\,0}) g(z_n,\bar t^{\,n-1}) }} \  \lambda_{n} (z)\ 
 \BB (\bar w)\, ,
    \end{split}
\end{equation}
where $z_1,z_n$ are defined in \eqref{zs}. 
This   proves the proposition~\ref{achimo}. \qed

\section{Conclusion}

This paper is a continuation of our work \cite{LPRS19a} to investigate the $\mathfrak{o}_{2n+1}$-invariant 
quantum integrable models which are defined by the $(2n+1)\times(2n+1)$ monodromy matrices satisfying 
the commutation relations \r{rtt} with $\mathfrak{o}_{2n+1}$-invariant $R$-matrix \r{Rmat}. 
To describe the space of states in these models we are using the method introduced in \cite{EKhP07} and developed in 
\cite{HutLPRS17a,HutLPRS20} for the supersymmetric integrable models associated with 
Yangian double $DY(\mathfrak{gl}(m|n))$. In these papers the action of the upper-triangular monodromy matrix elements 
were used to describe the recurrent relations for the corresponding off-shell 
Bethe vectors and the action of the lower-triangular 
matrix elements were exploited to find recurrence relations for the higher coefficients in the summation formula for the
Bethe vectors scalar products. 

Analogous program in case of  $\mathfrak{o}_{2n+1}$-invariant integrable models is far from completion. 
The arguments of the paper \cite{HutLPRS20} cannot be directly repeated since we yet do not have clear picture 
on the structure of the summation formulas for the scalar products and properties of the higher coefficients. 
We hope to investigate this picture in our forthcoming publications using the identification of the higher coefficients
with the kernels in the integral presentation of the Bethe vectors through the currents \cite{BPR10}.

\section*{Acknowledgments}

This work was carried out in Skolkovo Institute of Science and Technology
under financial support of Russian Science Foundation within grant 19-11-00275.
Authors thank E.~Ragoucy and N.~A.~Slavnov for fruitful discussions.

\end{document}